%% file: sample-sigconf.tex
\documentclass[sigconf]{acmart}
\usepackage{booktabs} 
\usepackage{subfigure}
\usepackage{xspace}
\usepackage{threeparttable}
\usepackage{multirow}
\usepackage{stmaryrd}
\newcommand{\paratitle}[1]{\vspace{1.5ex}\noindent\textbf{#1}}
\newcommand{\ie}{\emph{i.e.,}\xspace}

\newcommand{\eg}{\emph{e.g.,}\xspace}

\newcommand{\ignore}[1]{}
\usepackage{amsmath}
\usepackage{amssymb}

\setcopyright{rightsretained}

\ignore{
\acmDOI{10.475/123_4}

\acmISBN{123-4567-24-567/08/06}

\acmConference[]{ACM}{2019}{}
\acmYear{2019}
\copyrightyear{2019}

\acmArticle{4}
\acmPrice{15.00}
\editor{Jennifer B. Sartor}
\editor{Theo D'Hondt}
\editor{Wolfgang De Meuter}
}

\begin{document}
\title{A Long-Short Demands-Aware Model for Next-Item Recommendation}

\author{Ting Bai}
\affiliation{\institution{Renmin University of China}}
\email{baiting@ruc.edu.cn}

\author{Du Pan}
\affiliation{\institution{University of Montrea}}
\email{pandu@iro.umontreal.ca}

\author{Wayne Xin Zhao}
\affiliation{\institution{Renmin University of China}}
\email{batmanfly@ruc.edu.cn}

\author{Ji-Rong Wen}
\affiliation{\institution{Renmin University of China}}
\email{jirong.wen@gmail.com}

\author{Jian-Yun Nie}
\affiliation{\institution{University of Montreal}}
\email{nie@iro.umontreal.ca}


\begin{abstract}
Recommending the right products is the central problem in recommender systems, but the right products should also be recommended at the right time to meet the demands of users, so as to maximize their values. 
Users' demands, implying strong purchase intents, can be the most useful way to promote products sales if well utilized.
Previous recommendation models mainly focused on user's general interests to find the right products. However, the aspect of meeting users' demands at the right time has been much less explored. 
To address this problem, we propose a novel Long-Short Demands-aware Model (LSDM), in which both user's interests towards items and user's demands over time are incorporated. 
We summarize two aspects: termed as long-time demands (\eg purchasing the same product repetitively showing a long-time persistent interest) and short-time demands (\eg co-purchase like buying paintbrushes after pigments).
To utilize such long-short demands of users,
we create different clusters to group the successive product purchases together according to different time spans, and use recurrent neural networks to model each sequence of clusters at a time scale. The long-short purchase demands with multi-time scales are finally aggregated by joint learning strategies.
Experimental results on three real-world commerce datasets demonstrate the effectiveness of our model for next-item recommendation, showing the usefulness of modeling users' long-short purchase demands of items with multi-time scales.
\end{abstract}

%
\ignore{

\ccsdesc[500]{Computer systems organization~Embedded systems}
\ccsdesc[300]{Computer systems organization~Redundancy}
\ccsdesc{Computer systems organization~Robotics}
\ccsdesc[100]{Networks~Network reliability}
}

\keywords{Long-Short Purchase Demands, Multi-time Scales, Next-Item Recommendation}

\maketitle

\input{sec-intro}

\input{sec-model}

\input{sec-experiment}

\input{sec-related}

\input{sec-conclusion}

\bibliographystyle{ACM-Reference-Format}
\bibliography{bai}


\end{document}

%% file: sec-intro.tex
\section{Introduction}
Purchase demand refers to a user's desire and willingness to pay a price for a specific product.
It implies a user's strong purchase intent for a product, and can be the most useful way to promote products sales if well utilized ~\cite{ha2002customer,skaer1993effect}. 
Users' purchase demands are time sensitive~\cite{afeche2015pricing}, hence a good recommender system should not only be able to find the right products, but also recommend them at the right time to meet the demands of users, so as to maximize their values. 
So far, the majority of recommendation models, \eg collaborative filtering~\cite{he2016fast,koren2008factorization} and sequence-based models~\cite{wang2015learning,li2017neural,quadrana2017personalizing}, have mainly focused on modeling user's general interests to find the right products, while the aspect of meeting users' demands at the right time has been much less explored. Based on this consideration, our aim is to learn the time-sensitive purchase demands from users' purchase history, and to leverage such information to better understand the real-time demand of a user, so as to make a more accurate prediction of the next item. 
To fulfill the above purpose, we have to address two challenging issues: (1) \emph{ how to characterize} the time sensitive purchase demands of users; (2) \emph {how to utilize} such purchase demands in our model for predicting next-item of users.

To characterize the time sensitive demands, inspired by the studies in marketing strategies and human behaviors~\cite{tsai2004purchase,nowak1998dynamical,rahimi2013location}, we summarize two aspects of the time sensitive demands: termed as \emph{long-time demands} and \emph{short-time demands}.
Long-time demands refer that a user purchase the same product repetitively, showing a long-time persistent interest~\cite{bhagat2018buy}; while short-time demands refer the co-purchase of items~\cite{guidotti2017next}, \eg buying paintbrushes after pigments.
These two kinds of demands are commonly seen and had been much accounted in marketing~\cite{tsai2004purchase}. For example, online companies \eg contact lenses company Clearly\footnote{https://www.clearly.ca/} and drug mart shoppers\footnote{https://www1.shoppersdrugmart.ca/en/health-and-pharmacy/patient-contact} predict the long-time repeated purchase demands of users by estimating the service life of products user had purchased. The companies will send emails or notifications to users after authorizations, with the aim of reminding users to refill up products before they run out of the products. 
The short-time demands are also identified as a marketing strategy for promotional activities in online retail shops~\cite{ha2002customer}, \eg the system will recommend some co-purchased products after you had bought one related products~\cite{guidotti2017next,yap2012effective}. 
Although such long- and short-time demands of users can be estimated by some simple quantity or statistics scenarios, in most e-commerce websites, the service life of a large amount of products are hard to estimated, and co-purchase items may also vary due to different using or purchasing habits of users.

For utilizing such purchase demands, we propose a novel Long-Short Demands-aware Model (LSDM), in which both user's interests towards items and user's demands over time are incorporated.
\begin{figure*}
  \centering
    \includegraphics[width=0.8\linewidth]{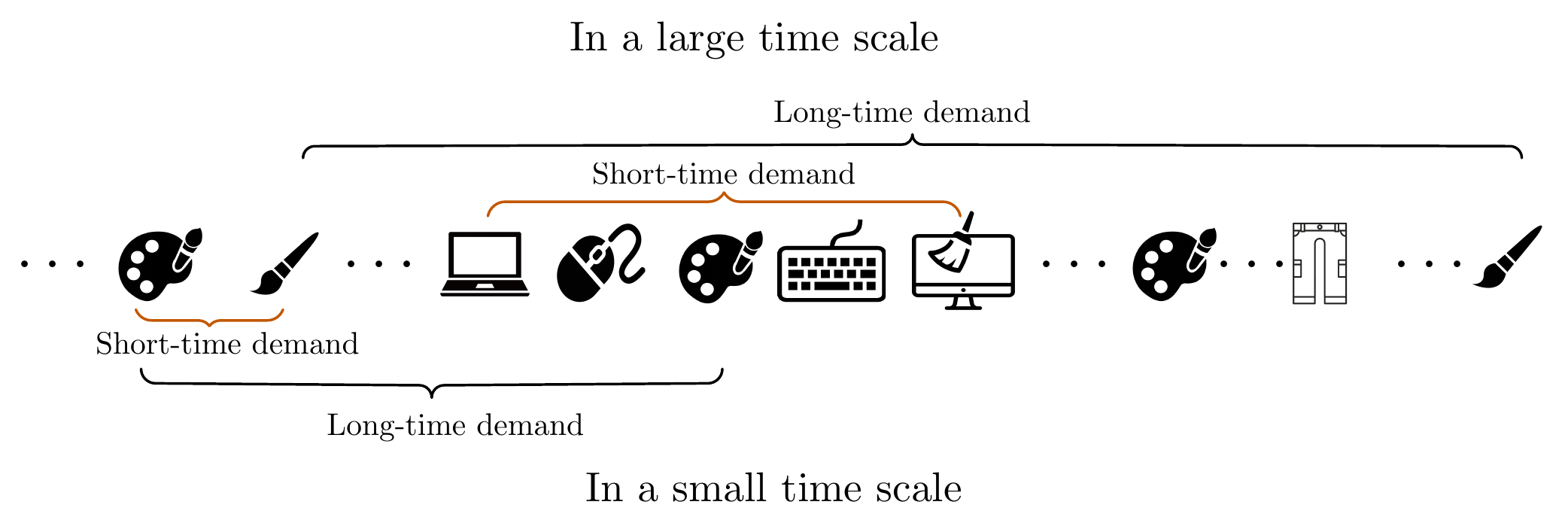}
  \caption{An illustration of long-short purchase demands in different time scales. We see that the user buys pigments and brushes repetitively (long-time demands) but with different time frequencies, and that the co-purchase (short-time demand) of screen cleaner is due to the purchase of a laptop computer quite some time earlier, while the co-purchase of brush occurs right after the purchase of pigment at the beginning of the sequence.}
  \label{fig:model} 
\end{figure*}
To model users' long-short time demands, 
We use a time scale (detail see in Sec.~\ref{sec-demands}) to cluster the successive product purchases within a time span together (\ie short-time demands) in order to cope with users' repeated purchase demands of items at a certain time frequency (\ie long-time demands).
However, the repeated purchase frequencies of the same or similar products could vary greatly, \eg a user may purchase milk more frequently than the detergent. Such different purchase demands may not be easily discovered from a single time scale which did in most session-based models~\cite{hidasi2015session,li2017neural,jannach2017recurrent}. Also, the single time scale may be insufficient to group the various types of the co-purchase items. In our model, we design multiple-time scales to learn from different types of long-short purchase demands. More specifically, we propose a hierarchical neural architecture with multi-time scales in LSDM (see in Figure~\ref{fig:model}), in which users' purchase preferences of items over time are captured by recurrent neural networks, and long-short purchase demands in different time scales are address by joint learning strategies.

We found limited research about demands-aware purchase recommendations and none of that deals with the real-world purchase data on e-commerce websites.
Our contributions in this paper are as follows:
\begin{itemize}
\item We propose a novel Long-Short Demands-aware Model (LSDM), in which both users' interests towards items and user demands over time are incorporated. To the best of our knowledge, such end-to-end neural model which automatically incorporate users' purchase demands into recommendation model has been much less explored.
\item We characterize two aspects of users' purchase demands: long-time demands (\ie repeated purchase demands) and short-time demands (\ie successive purchase demands), and incorporate them into a hierarchical architecture with multi-time scales.
\item Experimental results on three real-world commercial datasets (Ta-Feng, BeiRen and Amazon) demonstrate the effectiveness of our model for next-item recommendation, showing the usefulness of modeling user long-short purchase demands of items.
\end{itemize}

The rest of this paper is organized as follows: we first introduce some preliminary definition in Section 2. Following it, section 3 presents our long-short demands-aware model for recommendation and section 4 introduces the experiments. Related work is summarized in section 5. Finally, section 6 concludes this paper.

%% file: sec-model.tex
\section{Preliminaries}
In this section, we formulate the next-item recommendation  problem and give a detail explanation of some key concepts (\eg long-short purchase demands and time scales) in our paper.

\subsection{Problem Statement}
Assume we have a set of users and items, denoted by $U$ and $I$ respectively. Let $u\in U$ denote a user and $i\in I$  an item. The number of users and items are denoted as $|U|$ and $|I|$ respectively.
Given a user $u$, his purchase records $I^{u} $ is a sequence of items sorted by time, which can be represented as $I^{u}=\{i^{u}_{t_1}, i^{u}_{t_2}, ..., i^{u}_{t_j},...,i^{u}_{t_n}\}$, where  $i^{u}_{t_j}\in I$ is the item purchased by user $u$  at time $t_j$. 
Given the purchase history $I^{u}$ of a user $u$, the next-item recommendation aims to predict the next item that $u$ would probably buy at time $t_{n+1}$:
\begin{align}
P(i^{u}_{t_{n+1}})=\mathcal{F}(i \in I|u,I^{u}),
\label{prediction}
\end{align}
where $P(i^{u}_{t_{n+1}})$ is the probability of item $i \in I$ being purchased by $u$ at the next time $t_{n+1}$, and $\mathcal{F}$ is the prediction function. The prediction problem can also be formulated as a ranking problem of all items for each user. With the ranked list of all items, we recommend the top $K$ items to the user.

\subsection{Long-Short Purchase Demands}\label{sec-demands}
Purchase demands refer to a user's desire and willingness to pay a price for a specific product. Inspired by the studies in marketing strategies and human behaviors~\cite{tsai2004purchase,nowak1998dynamical,rahimi2013location}, we summarize two aspects of the time sensitive demands: termed as long-time demands and short-time demands.

\paratitle{Long-time demands}. Long-time demands refer to the fact that a user purchases the same product repetitively, showing a long-time persistent interest~\cite{bhagat2018buy}.

\paratitle{Short-time demands}. Short-time demands refer to the co-purchase purchase demands of users, \eg buying paintbrushes after pigments.

\paratitle{Time scale}. Time scale refers to a specific unit of time that divides up the purchase history of users into meaningful periods. We formulate the long-short demands with different time scales as follows:
given a user $u$ and his purchase records $I^{u}=\{i^{u}_{t_1}, i^{u}_{t_2}, ..., i^{u}_{t_j},...,i^{u}_{t_n}\}$, we use time window $T$ to define the time scale interval.
With all purchases within a time window being grouped together, the purchase records of a user can be grouped into sub-sequences by time scale $T$ as follows:
\begin{align}
I^{u}=\{I^{u}_{T_1}, I^{u}_{T_2},...,I^{u}_{T_j},...,I^{u}_{T_n}\}
\label{eq:timescale}
\end{align}
where $I^{u}_{T_j}=\{i^{u}_{t_j},...,i^{u}_{t_k}\}$ is a set of items within the same time window.

\paratitle{Multi-time scales}. Multi-time scales are composed with multiple different time scales.
As we mentioned above, for modeling users' long-short time demands, we use time scale to cluster the successive products together (\ie short-time demands), and cope with users' repeated purchase demands of items at different time frequencies (\ie long-time demands).
However, as shown in Fig.~\ref{fig:model}, the repeated purchase frequencies of same products could vary greatly, \eg user purchase pigments more frequently than the paintbrushes, which may not be easily discovered from a single time scale. Also, the single time scale may be insufficient to group the various types of the co-purchase items, \ie successively purchasing paintbrushes after pigments ( a kind of short-time demands in a small time scale) and a laptop and its cleaner are purchased separated by a few products ( a short-time demand in a slightly larger time scale).
Hence we use multi-time scales to model a more general long-short purchase demands (see Sec.~\ref{sec:multi-scales}). 

By organizing the purchase history of a user with multi-time scales, our model is powerful to model more general successive purchase demands (\ie short-time demands) and repeated purchase demands (\ie long-time demands).
The utility of multiple time scales to observe user's purchase sequence is well documented in studies in marketing strategies and human behaviors~\cite{tsai2004purchase,nowak1998dynamical,rahimi2013location}, which showed abundant evidence that human activities are largely regulated at several time scales and the final decision is based on interposition of them. The design of our model aims to fit these general observations on human behaviors.

\section{Our Proposed Model}\label{sec:model}
\begin{figure}
  \centering
    \includegraphics[width=0.95\linewidth]{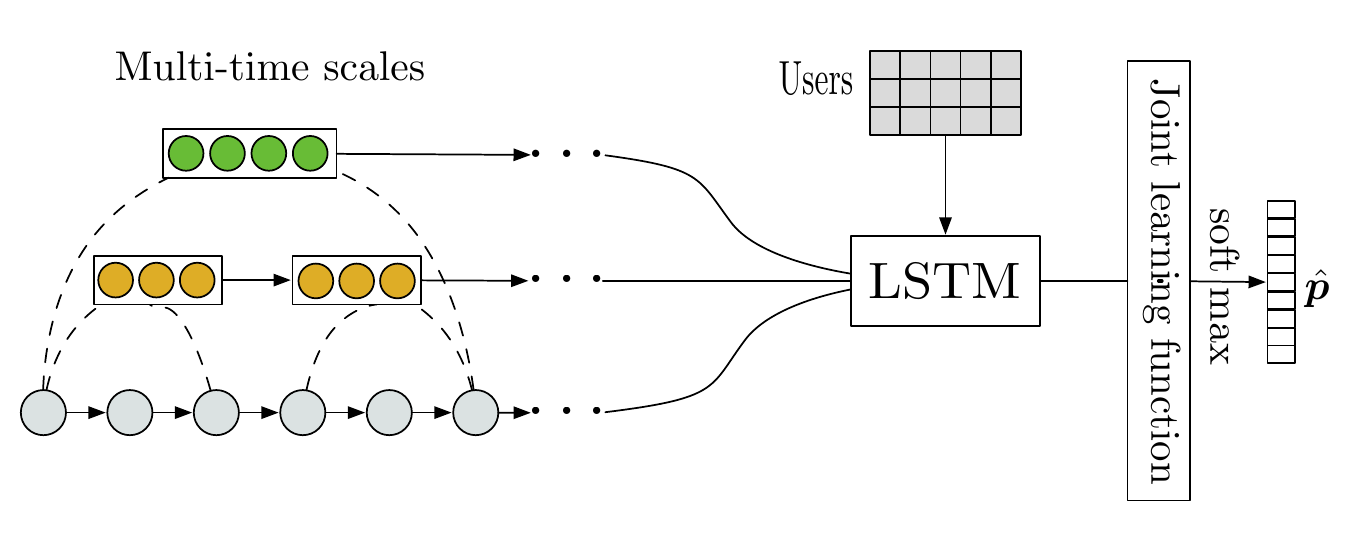}
  \caption{An overview architecture of LSDM.}
  \label{fig:model1} 
\end{figure}

In this section, we describe our Long-Short Demands-aware Model (LSDM) for personalized next-item recommendation.
we design a hierarchical neural architecture in LSDM with multi-time scales (see in Figure~\ref{fig:model1}). 
We use a sequential modeling of  users' purchase preferences of items over time via LSTM in each time scale. The final prediction of next-item is calculated by applying joint learning of the long-short purchase demands with multi-time scales. 
\subsection{Modeling Purchase Records}
Two main elements should be modeled: the sequence of purchased items and the user's interests.
\subsubsection{Encoding Attentive Transaction Information.}
In each time scale, we use a time window $T$ (see in Eq.~\ref{eq:timescale}) to group all purchases within a time window together, which is termed as a transaction. Given a user $u$ and his purchase records $I^{u}=\{I^{u}_{T_1}, I^{u}_{T_2},...,I^{u}_{T_j},...,I^{u}_{T_n}\}$, 
a transaction at step $T_j$ is $I^{u}_{T_j} \in I^{u}$. We represent the information of the set of items $I_{T_j}$ using an $|I|$-dimensional one-hot representation, denoted by $\mathbf{e}^{I}_{T_j}\in \mathbb{R}^{|I|\times 1}$, in which only the entry corresponding to the item involved in the transaction $I_{T_j}$ will be set to 1. Then a lookup layer is applied to translate each item $i_{T_j} \in I_{T_j}$ into a latent vector
\begin{align}
\mathbf{v}^{i}_{T_j}=&\textsc{lookup}(\mathbf{P}^{\top}, \mathbf{e}^{i}_{T_j}),
\label{latent_vector}
\end{align}
where $\mathbf{P}\in \mathbb{R}^{D \times 1}$ is the transformation vector for lookup, and $D$ is the embedding dimension of each item.
To obtain the representation of $I_{T_j}$, we adopt a concatenation operation to integrate the information of all items in transaction $I_{T_j}$ by
\begin{align}
\mathbf{v}^{I}_{T_j}=&\textsc{concat}(\mathbf{v}^{i}_{T_j}| i_{T_j} \in I_{T_j}),
\end{align}
where $\mathbf{v}^{I}_{T_j} \in \mathbb{R}^{D \times |I_{T_j}|}$ is the latent vector of transaction $I_{T_j}$. $|I_{T_j}|$ is the number of items in $I_{T_j}$. Since the number of items in each transaction varies, we use a masked zero-padding value in the embedding layer to convert each transaction to a fixed-dimension of representation vector.

Inspired by the success of attention mechanism in capturing the important information of previous states~\cite{luong2015effective,li2017neural}, we adopt an attention mechanism to relay user's appetite of a transaction to another. For a transaction $I_{T_j} \in I^{u}$, we assume the user's appetite for the transaction is $\mathbf{a}^{I}_{T_j}$. $\mathbf{a}^{I}_{T_j} \in \mathbb{R}^{D \times |I_{T_j}|}$ is initialized randomly and learned through the training process over all transactions.
We integrate attention weights with the latent vector of the transaction $\mathbf{v}^{I}_{T_j}$ as follows
\begin{align}
\mathbf{\tilde{v}}^{I}_{T_j}=&(\mathbf{v}^{I}_{T_j} \odot \mathbf{a}^{I}_{T_j}|I_{T_j}\in I^{u}),
\label{attentive-trans}
\end{align}
where ``$\odot$'' denotes the element-wise product of two vectors.
Once we have obtained the attentive representations for each transaction $\mathbf{\tilde{v}}^{I}_{T}$, we introduce how to model user's sequential behavior.
\subsubsection{Modeling User's Sequential Behavior.}
Given a user $u$, we represent it using a $|U|$-dimensional one-hot representation, denoted by $\mathbf{e}_{u}\in \mathbb{R}^{|U|\times 1}$. Then we apply a lookup layer to transform the one-hot vectors of $u$ into latent vectors
\begin{align}
\mathbf{v}_{u}=\textsc{lookup}(\mathbf{W}^{\top}, \mathbf{e}_{u}),
\label{uvector}
\end{align}
where $\mathbf{W}\in \mathbb{R}^{D \times 1}$ is the transformation vector for lookup.

The sequence of transactions at time granularity $T$ of user $u$ is $I^{u}=\{I^{u}_{T_1}, I^{u}_{T_2}, ..., I^{u}_{T_N}\}$.  We obtain the representation of each attentive transaction $\tilde{\mathbf{v}}^{I}_{T_j}$  in Eq.~\ref{attentive-trans}. The sequence of all attentive transactions can be represented as $\tilde{\mathbf{v}}^{I}=\{\tilde{\mathbf{v}}^{I}_{T_1},\tilde{\mathbf{v}}^{I}_{T_2},...,\tilde{\mathbf{v}}^{I}_{T_N}\}$.
To model the sequential behavior of a user, we adopt Long Short-Term Memory (LSTM) networks~\cite{hochreiter1997long}, which have proven effective in solving sequence learning problems~\cite{he2017translation}.
The input of LSTM at step $T_j$ is  $\tilde{\mathbf{v}}^{I}_{T_j}$. The output of LSTM (\ie hidden state) is represented as $\mathbf{v}^{h}_{T_j} \in \mathbb{R}^{D \times 1}$.
We model the interaction of user $u$ and  transaction $I_{T_j}$ by
\begin{align}
\tilde{\mathbf{v}}^{h}_{T_j}= \mathbf{v}_{u} \odot \mathbf{v}^{h}_{T_j},
\label {interaction}
\end{align}
where $\mathbf{v}_{u} \in \mathbb{R}^{D \times 1}$ (see in Eq.~\ref{uvector}) is the embedding vector of user $u$ and $\tilde{\mathbf{v}}^{h}_{T_j} \in \mathbb{R}^{D \times 1}$ is the interaction vector of user and current transaction. By updating user's latent vector $ \mathbf{v}_{u}$ at each step of transaction, our model can learn user's evolving interests to  items in the sequential records.

Given a user $u$ and his or her previous transactions $I^{u}_{T_1,T_N}$, we define the probability of an item $i$ being purchased in the next transaction $I^{u}_{T_{N+1}}$ by a softmax function

\begin{align}
p_i=p(i \in I^{u}_{T_{N+1}}|u, I^{u}_{T_1,T_N})= \frac {\exp ( \tilde{\mathbf{v}}^h_{T_N} \cdot \mathbf{v}_{i}^{\top})}
{\sum^{|I|}_{j=1}\exp(\tilde{\mathbf{v}}^h_{T_N} \cdot \mathbf{v}_{j}^{\top})},
\label {prob}
\end{align}
where $\tilde{\mathbf{v}}^h_{T_N}\in \mathbb{R}^{D \times 1}$ is the interaction vector of the user and the transaction at step $T_N$ (see in Eq.~\ref{interaction}). For the whole set of items $i \in I$, the predicted probability can be represented as $\mathbf{p}=\{p_{i_1}, p_{i_2},...,p_{i_{|I|}}\}$.
\subsection{Joint Learning with Multi-Time Scales}\label{sec:multi-scales}
As introduced in Sec.~\ref{sec-demands},  we use the time scales with different time windows $T$ to capture the long and short purchase demands of users. 
The different time scales can be denoted as  $\mathcal{T}=\{ T^{1}, T^{2},...,T^{c}\}$.
At each training epoch, the prediction results (see in Eq.~\ref{prob}) of the next step at all time scales are denoted as $\mathbf{p}_{\mathcal{T}}=\{\mathbf{p}_{T^{1}}, \mathbf{p}_{T^{2}},...,\mathbf{p}_{T^{c}}\}$, where $\mathbf{p}_{\mathcal{T}} \in \mathbb{R}^{|\mathcal{T} \times |I|}$ is the matrix of prediction results of  our long-short demands-aware model. Then we feed $\mathbf{p}_{\mathcal{T}}$ into a joint learning function to generate the final prediction results, denoted as
\begin{eqnarray}
\tilde{\mathbf{p}}= \mathcal{S}(\mathbf{p}_{\mathcal{T}}),
\label {joint-prob}
\end{eqnarray}
where $\tilde{\mathbf{p}}=\{\tilde{p}_1,\tilde{p}_1,...,\tilde{p}_{|I|}\}$ is the final prediction results of our model and $\mathcal{S}$ is the joint learning function.

The joint learning function $\mathcal{S}$ is flexible to be extended to an arbitrarily complex method. We will discuss it in the experiment section (see in $\mathbf{Q3}$).
In our experiments, we consider two kinds of joint learning functions: linear (\ie average, max and weighted joint learning) and non-linear (\ie multilayer perceptron joint learning) functions. Given the set of time scales $\mathcal{T}=\{T^{1}, T^{2},...,T^{c}\}$ and their corresponding prediction results $\mathbf{p}_{\mathcal{T}}=\{\mathbf{p}_{T^{1}}, \mathbf{p}_{T^{2}},...,\mathbf{p}_{T^{c}}\}$, the four joint learning functions are defined as follows
\begin{itemize}
\item Average joint learning function.
It uses the average of the prediction results of all time scales.
\begin{eqnarray}
\tilde{\mathbf{p}}= \frac{\sum_{c=1}^{|\mathcal{T}|}(\mathbf{p}_{T^c} )}{|\mathcal{T}|}.
\label {avg-prob}
\end{eqnarray}

\item Max joint learning function. It uses a maximum operation on the prediction results of all time scales.
\begin{eqnarray}
\tilde{\mathbf{p}}= \max_{c=1}^{|\mathcal{T}|}(\mathbf{p}_{T^c}).
\label {max-prob}
\end{eqnarray}

\item Weighted joint learning function.  It learns user's preference of all items in the different time scales by the weights. Given a time granularity $T^{c} \in \mathcal{T}$, we learn a weight vector $\mathbf{a}_{c} \in \mathbb{R}^{|I|\times 1} $ of $T^{c}$. $\mathbf{a}_{c}$ is initialized randomly and learned automatically in the training process of our model. We obtain the weighted prediction results of all time scales by
\begin{eqnarray}
\tilde{\mathbf{p}}=\frac{\sum_{c=1}^{|\mathcal{T}|}(\mathbf{a}_{c} \odot \mathbf{p}_{T_{c}})}{|\mathcal{T}|}.
\label {weight-final-prob}
\end{eqnarray}

\item Multilayer Perceptron (MLP) joint learning function. This is a non-linear joint learning function. We examine if this function is more powerful to capture user's preference at different time scales. We first concatenate the prediction results at different time scales by
\begin{align}
\mathbf{P}&=\textsc{concat}(\mathbf{p}_{T^{1}}, \mathbf{p}_{T^{2}},...,\mathbf{p}_{T^{c}}),
\label {concat-prob}
\end{align}
where $\textsc{concat}$ is the concatenate operation of vectors.
Then a multilayer perceptron~\cite{gardner1998artificial} is used to obtain $\tilde{\mathbf{p}}$ by
\begin{align}
 \phi_{1}(\textbf{z}_1)&= \textbf{f}_1(\textbf{W}_1^{\top}\mathbf{P}+\textbf{b}_1),\\\nonumber
&...... \\ \nonumber
\phi_{L}(\textbf{z}_L)&=  \textbf{f}_L(\textbf{W}_L^{\top}\mathbf{P}+\textbf{b}_L),\\\nonumber
\tilde{\mathbf{p}}& =\sigma(\textbf{h}^{\top}\phi_{L}(\textbf{z}_L)),
\label {mlp-prob}
\end{align}
where $\textbf{W}_x$, $\textbf{b}_x$, and $\textbf{f}_x$ denote the weight matrix, bias vector, and activation function for the $x$-th layer's perceptron respectively. 
For activation functions $\sigma$ of MLP layers, one can freely choose among sigmoid, hyperbolic tangent (tanh), and Rectifier (ReLU), among others.
\end{itemize}

\subsection{The Loss Function for Optimization}
Our LSDM is optimized by a joint learning process. The objective functions to be optimized in all time scales  $\mathcal{T}=\{T^1, T^2,...,T^{c}\}$ is denoted as $L_{\mathcal{T}}=\{L_{T^1}, L_{T^2},..., L_{T^{c}}\}$. The objective function in our LSDM can be defined as
\begin{eqnarray}
L_{\mathcal{T}} &= & \min (\mathcal{S}(L_{T^1}, L_{T^2},..., L_{T^{c}}|\Theta_{\mathcal{T}})\\ \nonumber
 &= &  \min (\mathcal{S}(\min (\mathcal{R}(I_{T^{1}}|\Theta_{T^1})),\\ \nonumber
 & & \min (\mathcal{R}(I_{T^2}|\Theta_{T^2})),..., \min (\mathcal{R}(I_{T^{c}}|\Theta_{T^{c}})))),  \nonumber
\end{eqnarray}
where $\{I_{T^1}, I_{T^2},..., I_{T^{c}}\}$ are sequences at different time scales. The model parameters on different time scales are $\{\Theta_{T^1}, \Theta_{T^2},...,\Theta_{T^{c}}\}$. $\mathcal{S}$  (see in Eq.~\ref{joint-prob}) and $\mathcal{R}$ are the learning function of LSDM and each time scale model respectively. The parameters to be learned in LSDM are $[\Theta_{T^1}, \Theta_{T^2},..., \Theta_{T^{c}}, \Theta_{\mathcal{T}}]$.

For a time scale $T^{c}$, we adopt a weighted cross-entropy as the optimization objective at each step of LSTM:
\begin{align}
L_{T^{c}} = \sum\limits_{u \in U} \sum \limits_{I_{T_{j}} \in I^{u}} \sum \limits_{i \in I_{T_{j}}}(&  -m \cdot y_i \cdot \log {p}_i  \\ \nonumber
& -n \cdot (1-y_i) \cdot \log (1-{p}_i))
\label{object-function}
\end{align}
where $T_j$ is the $j$th transaction in time scale $T^{c}$. ${p}_i$ is the probability of an item $i$ being purchased in the next transaction (see in Eq. ~\ref{prob}). If an item $i$ is purchased in the the next transaction, $y_{i}=1$, otherwise, $y_{i}=0$. $m$ and $n$ are the weights of positive and negative instances (\ie item $i$ is purchased or not in the next transaction). These weights are used to cope with unbalanced number of positive and negative examples. In our experiments, the ratio of positive and negative instances is about 500, so we set $m$ to 500 times higher than $n$ to reduce the training bias.

In our LSDM, the objective function is defined as: 
\begin{equation}
L_{\mathcal{T}} = \sum \limits_{u \in U}\sum \limits_{i \in I_{T_N}}( -m \cdot y_i \cdot \log \tilde{p}_i  -n \cdot (1-y_i) \cdot \log(1-\tilde{p}_i)),
\label{mul-object}
\end{equation}
where $\tilde{p}_i$ is the probability of items in our LSDM (see Eq.~\ref{joint-prob}).

After training, given a user's history purchase records, we can obtain the probability of each item $i$ being purchased at the next step according to Eq.~\ref{prob}. We than rank the items according to their probability, and select top $K$ results as the final recommended items to the user.

%% file: sec-experiment.tex
\section{Experiments}
We conduct experiments on three real-world datasets to verify the effectiveness of our proposed model for next-item recommendation. In particular, we aim at answering the following questions:
\begin{itemize}
\item \textbf{Q1}: Are purchase demands of users really useful for next-item recommendation task?
\item \textbf{Q2}: Are multi-time scales more powerful to capture the long-short purchase demands of users?
\item \textbf{Q3}: Is the joint learning function effectively incorporate users' purchase demands with multi-time scales?

\end{itemize}
In the following section, we will first introduce our experimental settings, including datasets, baselines, and evaluation metrics. Then we will analyze the various experimental results to answer the three questions one by one.
\subsection{Experimental Settings}
\subsubsection{Datasets.}
We experiment with three real-world datasets: Ta-Feng\footnote{\begin{scriptsize}http://www.bigdatalab.ac.cn/benchmark/bm/dd?data=Ta-Feng\end{scriptsize}}, BeiRen\footnote{\begin{scriptsize}http://www.bigdatalab.ac.cn/benchmark/bm/dd?data=Ta-Feng\end{scriptsize}} and Amazon\footnote{\begin{scriptsize}http://jmcauley.ucsd.edu/data/amazon/\end{scriptsize}}. Ta-Feng and BeiRen are two online shopping datasets with real purchase records of users.
These two datasets are the only public available ones we are aware of that contain the real purchase history of users. We also conduct experiments on Amazon dataset, which is commonly used in many recommender models. Amazon a review dataset: users' purchase records are collected from reviews\footnote{Note that a review usually implies a purchase, but a user may purchase an item without leaving a review.}
\begin{itemize}
\item {Ta-Feng}~\cite{wang2015learning} is a grocery shopping dataset, it covers products from food, office supplies to furniture. We use the data in a quarter (\ie from December 2000 to February 2001) of shopping transactions of the Ta-Feng supermarket. Since it is unreliable to include users with few purchase times or limited active time for evaluation, we first remove the products which were bought less than 15 times and then keep users with purchase records in at least 5 weeks. We leave 7,044 items and 1,951 users with total 90,986 purchase records. The average number of purchase records of users are 50 and and the average times each item had been purchased is 14.
\item {BeiRen}~\cite{le2017basket} is an online shopping dataset. We use 4 months (April 2013 to July 2013) in BeiRen and conduct the same data filtering methods as conducted in Ta-Feng dataset. Finally, we obtain 211,519 purchase records involving 3,264 users on 5,818 items. The average number of purchase records of users is 65.
\item{Amazon}~\cite{he2016ups} is one of the largest Internet retailer in the world. We only can obtain the product review records. since users usually only post reviews after they made product purchases, we assume that reviews on Amazon correspond to actual purchases most of the time~\cite{bai2018characterizing}. We use review records in half a year (\ie from January 1st, 2014 to  June 30th, 2014). We first remove the products which have been purchased less than 5 times and then retain users with purchase records in at least 5 weeks. We obtain 6,092 items and 1,443 users with 15,811 purchases. The average number of product being reviewed by user is 11.
\end{itemize}
 
\subsubsection{Time Scales Selection.}
To implement the multi-time scale model, we need to determine what time scales to use. 
As introduced in Sec.~\ref{sec-demands},
time scale is used to cluster the successive products together (\ie short-time demands), and cope with users' repeated purchase demands of items at a certain time frequency (\ie long-time demands).
Inspired by the studies in marketing strategies and human behaviors~\cite{tsai2004purchase,nowak1998dynamical,rahimi2013location}, which showed abundant evidence that regularities structure is a defining feature of human activities, and the strongest influence is made by daily, followed by weekly and seasonal regular structures. It has also been found that ``rhythms of life" are superimposition of different regularities~\cite{nowak1998dynamical}.
So it is well justified to model user's long-short purchase demands by following these strongest rhythms, \ie daily, weekly and seasonal.
Considering the time periods in our datasets are less than half a year, in our experiments, 
we select the daily and weekly scales in our model, \ie  daily scale and weekly scale. We also keep the original sequence information of users' purchase history, we call it item scale.
The usefulness of these rhythm based time scales (\ie daily and weekly scales) are demonstrated in the following Sec.~\ref{sec:timescale}.

\subsubsection{Evaluation Metrics.}
Given a user, we infer the next item that the user would probably buy at next purchase. Each candidate method will produce an ordered list of items for the recommendation. We adopt two widely used ranking-based metrics to evaluate the performance of a ranked list: Hit ratio at rank $k$ (Hit@$k$) and Normalized Discounted Cumulative Gain at rank $k$ (NDCG@$k$).

\begin{itemize}
\item {Hit ratio at rank $k$ (HR@$k$).}
Given the predicted ordered list of items for a user, Hit@$k$ is defined as:
\begin{equation}
Hit@k=\sum_{c=1}^{k}\mathbb{I}(i_c, u),
\end{equation}

\item {Normalized Discounted Cumulative Gain at rank $k$ (NDCG@$k$).}
Given the predicted ordered list of items for a user, NDCG@$k$ is defined as:
\begin{equation}
NDCG@k=\sum_{c=1}^{k}\frac{\mathbb{I}(i_c,u)}{\log(c+1)}
\end{equation}
where $c$ is the position of items in the ranking list. $\mathbb{I}(i_c, u)$ returns 1 if $i_c$ was adopted by user $u$ in original dataset, and 0 otherwise.
\end{itemize}

Hit@$k$ intuitively measures whether the test item is present in the Top-$K$ List, and it accounts for the position of the hit by assigning higher scores to the hit with higher ranks.
We report the top $K$ (\ie $K=5$ and $K=10$) items in the ranking list as the recommended set.

\subsubsection{Baseline Methods Compared.}
We compare our model with the state-of-the-art methods from different types of recommendation approaches, including:

\begin{itemize}
\item Pop. It ranks the items according to their popularity measured by the number of being purchased. This is a widely used simple baseline.
\item BPR~\cite{rendle2009bpr}. It optimizes the MF model with a pairwise ranking loss. This is a state-of-the-art model for item recommendation, but the sequential information is ignored in this method.
\item FPMC~\cite{rendle2010factorizing}: It learns a transition matrix based on underlying Markov chains. Sequential behavior is modeled only between the adjacent transactions.
\item HRM~\cite{wang2015learning}: It employs a neural network to conduct a nonlinear operations to integrate the representation of customers and purchase history of items from the adjacent transactions.
\item RRN ~\cite{wu2017recurrent}: This is a representative approach that utilizes RNN to learn the dynamic representation of users and items in recommender systems. Our model with sequence of items can be regarded as equivalent to RRN model. 
\item NARM~\cite{li2017neural}: This is a state-of-the-art approach in personalized session-based recommendation with RNN models. It uses attention mechanism to determine the relatedness of the past purchases in the session for the next purchase. As our datasets do not have explicit information of sessions, we simulate sessions by the transactions within each day. 
\end{itemize}

The above methods cover different kinds of the approaches in recommender systems: BPR is a classical method among traditional recommendation approaches; FPMC and HRM are representative methods which utilize the adjacent sequential information. RNN and NARM are recent methods using the whole sequential information for recommendation. Table~\ref{tb-baseline} summarizes the properties of different methods.
Our LSDM is a demands-aware model. Short-time demands of users are model by the local sequence information of items within a transaction, and long-time demands are captured by the global information from the whole sequence of records.

\begin{table}
\small
\tabcolsep 0.03in
\centering
\caption{Properties of methods. P: personalized? D: deep neural network model?  L: local subsequence information? G: global whole sequence information?  D: demands aware?}
\label{tb-baseline}
\begin{tabular}{c c c c c c c c} \hline
~ & Pop & BPR  & FPMC & HRM & RNN & NARM & LSDM \\ \hline \hline
P& $\bigtimes$ & $\surd$  & $\surd$ & $\surd$ &$\surd$ &$\surd$ &$\surd$\\
D& $\bigtimes$ &  $\bigtimes$ &$\bigtimes$ & $\surd$  &$\surd$ &$\surd$ &$\surd$ \\
L & $\bigtimes$ & $\bigtimes$ & $\surd$& $\surd$ & $\surd$ &$\surd$ &$\surd$ \\
G & $\bigtimes$ & $\bigtimes$   & $\bigtimes$  & $\bigtimes$ &$\surd$  &$\surd$ &$\surd$ \\
D &$\bigtimes$  & $\bigtimes$  &$\bigtimes$ & $\bigtimes$&$\bigtimes$ &$\bigtimes$ &$\surd$ \\ \hline
\end{tabular}
\end{table}

Other sequential methods, \eg DREAM~\cite{yu2016dynamic}, TransRec~\cite{he2017translation}, user-based RNN~\cite{donkers2017sequential} and HRNN~\cite{quadrana2017personalizing}, are similar to our baseline methods, so they are not included in our comparison. For sequence methods with auxiliary information, \eg content-based neural model~\cite{suglia2017deep,beutel2018latent}, neural tensor factorization~\cite{wu2018neural} and pattern mining based model~\cite{yap2012effective,song2014personalized,guidotti2017next,quadrana2018sequence}, we also do not make comparisons due to additional information used in them.

\subsubsection{Parameter Settings.}
The hyper parameters of each method, with which we obtain the best prediction results, are listed below.
(1) BPR: the latent factors are 300, 300,200, the learning rates are 0.001, 0.001, 0.0005 in Ta-Feng, BeiRen and Amazon datasets respectively.
(2) FPMC: the latent factors are 32,32 and 16, the learning rates are 0.015, 0.01, 0.001 in the three datasets.
(3) HRM: the embedding size is 40, the learning rate is 0.005 and droprate is 0.5 in all datasets.
(4) RRN: The embedding size is 50, the learning rate is 0.001 in all datasets. Batch size is set to 100, 20, 100 respectively. 
(5) NARM: the embedding sizes are 25, 15, 20, with 25, 25, 20 hidden units. The learning rates are 0.0001, 0.0008, 0.0008 and batch sizes are 256, 640, 640 in the three datasets.
(6) MGASM: The embedding size is 50, the learning rate is 0.001 in all datasets. Batch sizes are 100, 20, 100 in the three datasets respectively.

For all the methods, we take the last item of each user as the predicting target,  the penultimate item as the validation data for model selection, and the remaining part in each sequence as the training data to optimize the model parameters.

\subsection{Performance Comparison (RQ1)}
We present the results of Hit@$k$ and NDCG@$k$, (\ie $k=5$ and $k=10$) on the next-item recommendation performance in Table~\ref{tb-result}. 

\begin{table*}[!htbp]
\small
\centering
\caption{Performance comparison of different methods on the next-item recommendation task.}\label{tb-result}
\newcommand{\tabincell}[2]{\begin{tabular}{@{}#1@{}}#2\end{tabular}}
\setlength{\tabcolsep}{1mm}{
\begin{tabular}{c|c c c c|c c c c|c c c c} \hline
Datasets& \multicolumn{4}{c|}{Ta-Feng}& \multicolumn{4}{c| }{BeiRen} & \multicolumn{4}{c}{Amazon}\\ \hline\hline
Models & Hit@5 & Hit@10 & NDCG@5 & NDCG@10 & Hit@5 & Hit@10 & NDCG@5 & NDCG@10 & Hit@5 & Hit@10 & NDCG@5 & NDCG@10  \\ \hline\hline
Pop&  0.0731& 0.0862 & 0.0573 & 0.0663 & 0.1743 &0.1982 & 0.1035 & 0.1109 & 0.0133 &0.0188& 0.0095 &0.0114\\
BPR &  0.0928 &0.1065 & 0.0698  & 0.0822 &0.1814 & 0.2114&0.1255 & 0.1367 &0.0172   & 0.0249 & 0.0105  & 0.0121\\
FPMC & 0.0945 &0.1100& 0.0772 &0.0829&0.1843& 0.2155& 0.1273& 0.1390 & 0.0174 & 0.0256& 0.0102& 0.0122\\ \hline
HRM& 0.0912 & 0.1143 & 0.0770& 0.0825& 0.1863 & 0.2001& 0.1082&0.1120& 0.0169& 0.0236& \textbf{0.0121}& 0.0134\\
RRN & 0.0984&0.1117& 0.0720 &0.0833&  0.1869& 0.2190&0.1317 &  0.1430 & 0.0163& 0.0245& 0.0114 & 0.0136  \\
NARM &0.1021&0.1186& 0.0789&0.0833 &0.1824 & 0.2053 &  0.1487 &  0.1518 & 0.0177 & 0.0261 &  0.0117 &  0.0144 \\  \hline
LSDM & \textbf{0.1194*} & \textbf{0.1281*}  & \textbf{0.0824*} & \textbf{0.0890*} & \textbf{0.2187*} &  \textbf{0.2290*}& \textbf{0.1617*}& \textbf{0.1646*} &\textbf{0.0182*} & \textbf{0.0265} & 0.0119 & \textbf{0.0147}\\ \hline
\end{tabular}}
\begin{tablenotes}
\centering
\footnotesize
\item Note: LSDM uses three typical time scales, \ie item, daily and weekly ( which discussed in Sec.~\ref{sec:timescale}) and MLP joint learning function.
\item $``*"$ indicates the statistically significant improvements (\ie two-sided $t$-test with  $p<0.05$ ) over the best baseline.
\end{tablenotes}
\end{table*}

We have the following observations:

(1) Pop is the weakest baseline in all datasets, since it is a non-personalized method. BPR performs better than Pop, but is not as good as FPMC, which uses adjacent sequential information of the transition cubes. This shows that the local adjacent sequential information is useful in predicting the next item.

(2) HRM and RRN perform better than BPR and FPMC that do not use neural network in Ta-Feng and BeiRen datasets. It indicates that neural network is capable of modeling complex interactions between user's general taste and their sequential behavior. RRN performs better than HRM, which may lie in that RRN model uses recurrent neural network to learn from the whole sequential data, while HRM only utilizes the adjacent sequential information.

(3) NARM is the state-of-the-art neural model for sequential prediction task, and performs the best among all of the baseline methods except on  Hit on BeiRen dataset. The attention mechanism enables NARM to attend to the most related purchases in the sessions and generate more accurate results. The effect is most visible on NDCG.

(4) Our LSDM with three typical time scales, \ie item, daily and weekly, and MLP joint learning function ( which will discussed in Sec.~\ref{sec:timescale} and Sec.~\ref{sec-joint}) performs the best in almost all datasets.
LSDM significantly outperforms all the baseline methods on Ta-Feng and BeiRen datasets. This indicates that the long-short purchase demands information used in our model is useful in predicting the real-time purchase demands of next item. Compared with RRN, LSDM not only utilizes global information from the whole sequence, but also captures some more complex local sequential information by grouping items into transactions. Compared with NARM, the multi-time scales of LSDM can adaptively learn from different repeated purchase demands and co-purchase items, which creating a more fine-grained model for users. 

(5) In Amazon dataset, although our LSDM improves the recommendation performance, the results only significantly in Hit@$5$. Similarly, sequence model HRM and RRN also loss advantages compared with BPR and FPMC. The reason is that Amazon dataset is a review dataset, in which the purchase records of a user are collected from users' reviews. Users will not write reviews after all purchases, hence such incomplete sequence may not be well learned by sequence models.

To further verify whether there exists repeated purchase of items at different time scales, we calculate the percentage of users who at least periodically purchase one item (\ie an item is purchased successively in at least half of all transactions). The  percentages are 20.4\%, 41.2\%, 5.3\% at time scale of days, while 43.4\%, 75.0\%, 10.7\% at weeks in Ta-Feng, BeiRen and Amazon datasets respectively. This indicates that many users have repeated purchase records in purchase datasets Ta-Feng and BeiRen, while the repeated purchases are far smaller in Amazon, due to it is a review datasets with incomplete real purchase records. 
Due to the incomplete sequence information in Amazon dataset, in the following results analysis, we only use real purchase datasets Ta-Feng and BeiRen to demonstrate the effectiveness of our model.


\subsection{Usefulness of Multi-Time Scales (RQ2)}\label{sec:timescale}

 \begin{figure}
  \centering
  \subfigure[Ta-Feng]{
 \begin{minipage}[c]{0.5\textwidth}
   \centering
    \includegraphics[width=2.8 in]{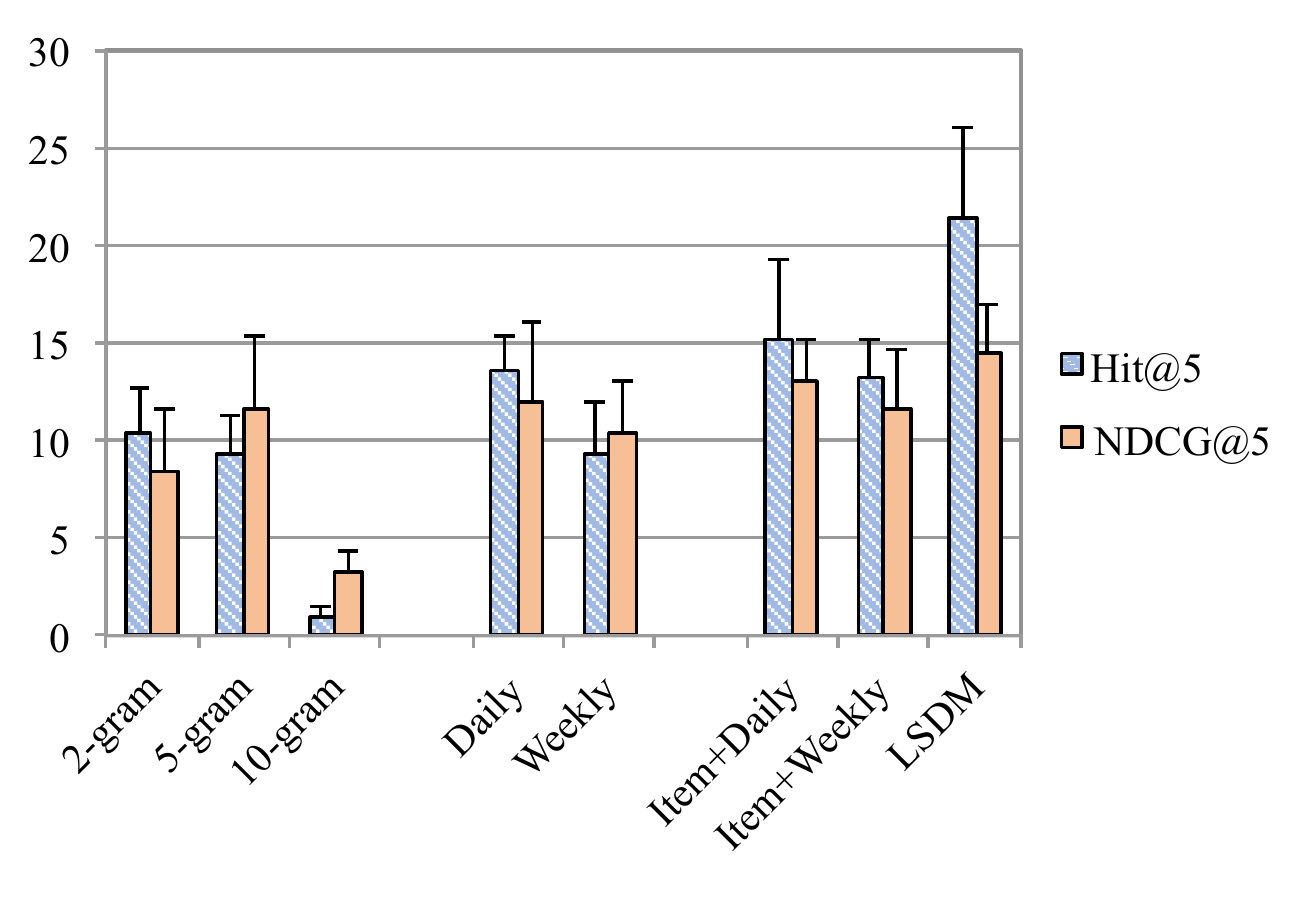}
    \end{minipage} 
    }
  \subfigure[BeiRen]{
   \begin{minipage}[c]{0.5\textwidth}
     \centering
    \includegraphics[width=2.8 in]{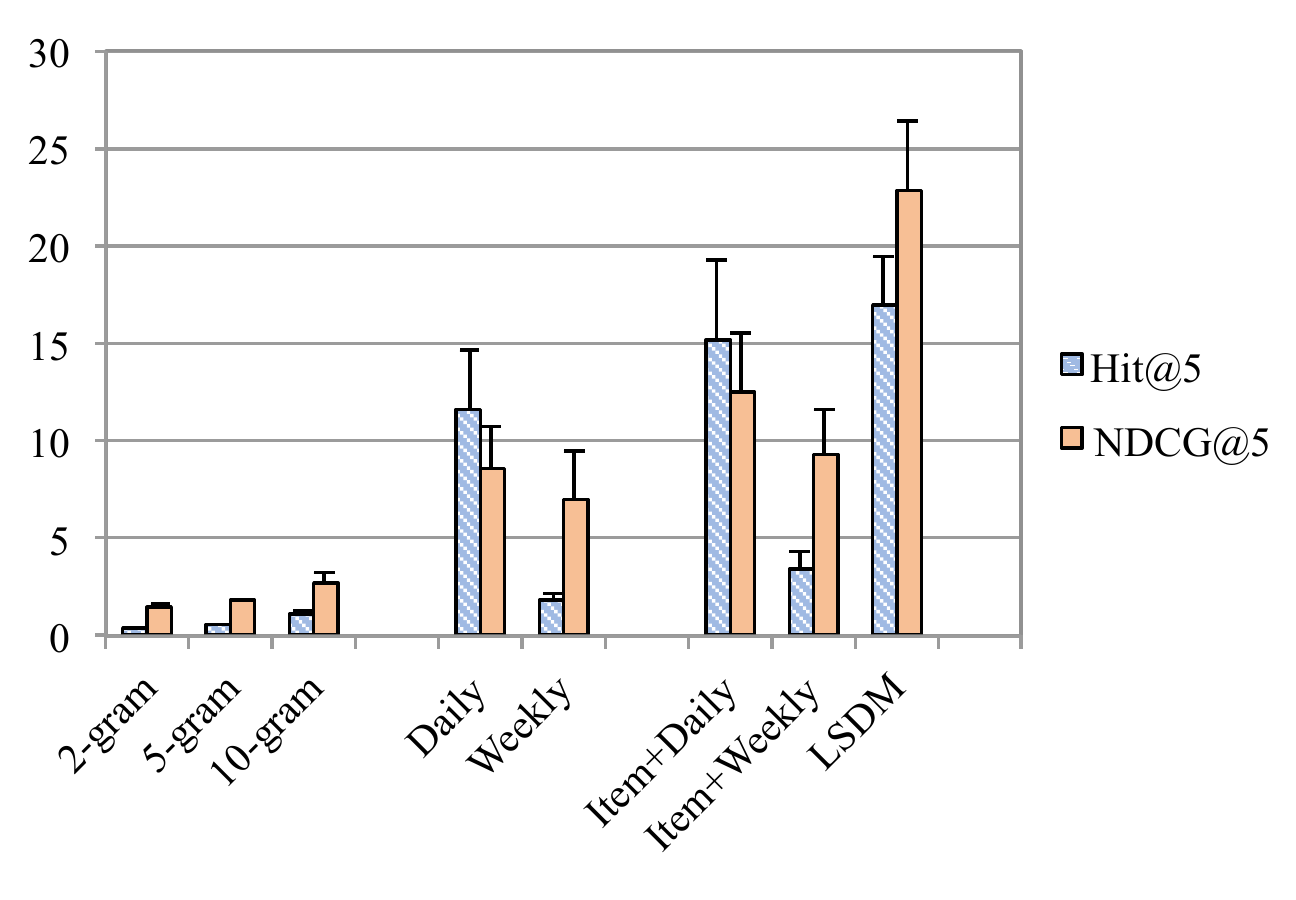}
        \end{minipage} }
  \caption{The increase percentage of no-rhythm, rhythm based and multi-time scales models compared to only item scale.}
  \label{fig-ngram} 
\end{figure}
In our LSDM, we use rhythm based time scales, \ie daily and weekly scales, and item scale. For demonstrate the usefulness of multi-time scales used in our model.
We address two issues:

\paratitle{The effectiveness of rhythm based time scales.} 
Following the ``rhythms of life" in society theory~\cite{nowak1998dynamical}, we use daily and weekly time scales (\ie LSDM$_{daily}$ and LSDM$_{weekly}$), termed as rhythm based time scale in our experiments.
To demonstrate the effectiveness of these rhythm based time scales, we generate other no-rhythm based time scales , \ie we cluster every two, five, ten items in the purchase sequence into a transaction. We borrow the terms in natural language processing, call these time scales as 2-gram, 5-gram and 10-gram scales (\ie LSDM$_{2-gram}$, LSDM$_{5-gram}$ and LSDM$_{10-gram}$). We compute the increased ratios on Hit@$k$ and NDCG@$k$ of the different time scales compared to the model with item scale only (\ie LSDM$_{item}$).

\paratitle{The usefulness of multi-time scales.} 
We further examine the usefulness of our multi-time scales, we compare our LSDM with degraded LSDM, \ie LSDM$_{item}$, LSDM$_{item+daily}$ and LSDM$_{item+weekly}$.


We compute the percentage of improvement on Hit@$5$ and NDCG@$5$ of the different methods over the model with item scale only (\ie LSDM$_{item}$), and present the results in Fig.~\ref{fig-ngram}.
We can see that: 
(1) It can be observed that both rhythm based LSDM and no-rhythm based LSDM methods are better than the model with item scale only. This shows that much of the underlying long and short purchase demands, \eg co-purchase information, may fail to be extracted from item sequence. Any time scales we consider is helpful to learn the long-short demands for next item prediction task;
(2) The rhythm based LSDM performs better than the no-rhythm based model on all datasets. This indicates that the time rhythms, \ie daily and weekly, are really help to capture user's purchase demands in the next item prediction;
(3) The performance of LSDM with multi-time scales is the best, \ie the performance of LSDM is better than degraded LSDM$_{item+daily}$ and LSDM$_{item+weekly}$. It indicates that user's complex purchase demands can be well captured by multi-time scales. The performance of daily scale is higher than that of weekly scale, which indicates that the time scale of a day is more useful than that of week on our datasets. However, this observation is highly dependent on the data. It is possible that larger time scale become more useful on a dataset of larger time span (\eg covering records of several years). Nevertheless, we can conclude now that different time scales are useful to detect different user purchase demands. These time scales tend to be complementary, leading to improved results when they are all considered.
The multi-time scales architecture in our model is flexible to learn from any time time scales, \eg the period years, which is easily to extend if we have a longer observation of the purchase records of users.

\subsection{Effects of Joint Training Strategy (RQ3)}\label{sec-joint}
We use a joint learning function $\mathcal{S}$ (see Eq.~\ref{joint-prob}) to integrate the purchase demands with multi-time scales in LSDM. In our experiments, we examine different joint learning functions: average, max, weighted and multilayer perceptron (MLP). The MLP method uses one layer, and sigmoid as activation function.
We present the results of our LSDM with the four joint learning functions on Hit@$k$ and NDCG@$k$, (\ie $k=5$ and $k=10$) in Table~\ref{tab-jointlearn}.
\begin{table}[!htbp]
\small
\centering
\caption{Comparison of different joint functions.}\label{tab-jointlearn}
\begin{tabular}{c|c c c c } \hline
Evaluation & Hit@5& Hit@10& NDCG@5& NDCG@10\\ \hline \hline
\multicolumn{5}{c}{Ta-Feng}\\ \hline
Max& 0.1148 & 0.1246 & 0.0817& 0.0874 \\
Avg & 0.1180 &  0.1275 & 0.0802 &  0.0883 \\
Weighted& 0.1071 & 0.1266& 0.0819 & 0.0872 \\
MLP & \textbf{0.1194} &  \textbf{0.1281} & \textbf{0.0824} &  \textbf{0.0890}\\ \hline\hline
\multicolumn{5}{c}{BeiRen}\\ \hline
Max& 0.2107 & 0.2285 & 0.1496& 0.1534 \\
Avg & 0.2118 & 0.2234 & 0.1598  & 0.1604 \\
Weighted & 0.1976&0.2242 & 0.1480&0.1504 \\
MLP &\textbf{0.2187} & \textbf{0.2290 }&\textbf{0.1617} & \textbf{0.1646} \\ \hline\hline
\end{tabular}
\end{table}

We can see that the MLP joint learning function performs the best on both datasets. This implies that non-linear function (\ie MLP) is more effective to capture the purchase demands information in different time scales.
To further examine the effects of joint learning, we present the training loss and experimental performance on Hit@5 and NDCG@5 of LSDM and degenerated model with a single time scale (\ie item, daily and weekly) in Fig.~\ref{fig-training}.  We can see that: (1) The training loss is smaller in LSDM than other degenerated models, and the item scale is the worst from this perspective. (2) As the iteration increases, LSDM tends to outperform the degenerated models and converges faster than others; (3) The performance of degenerated LSDM with time scale of day is better than week, and the item scale is the worst. This indicates that much of the purchase behavior may can not be observed from the item sequence, while it is easier by using a larger time scale of day.
\ignore{
\begin{figure*}[htbp]
 \centering
  \subfigure[Ta-Feng]{
     \includegraphics[width=1.53in]{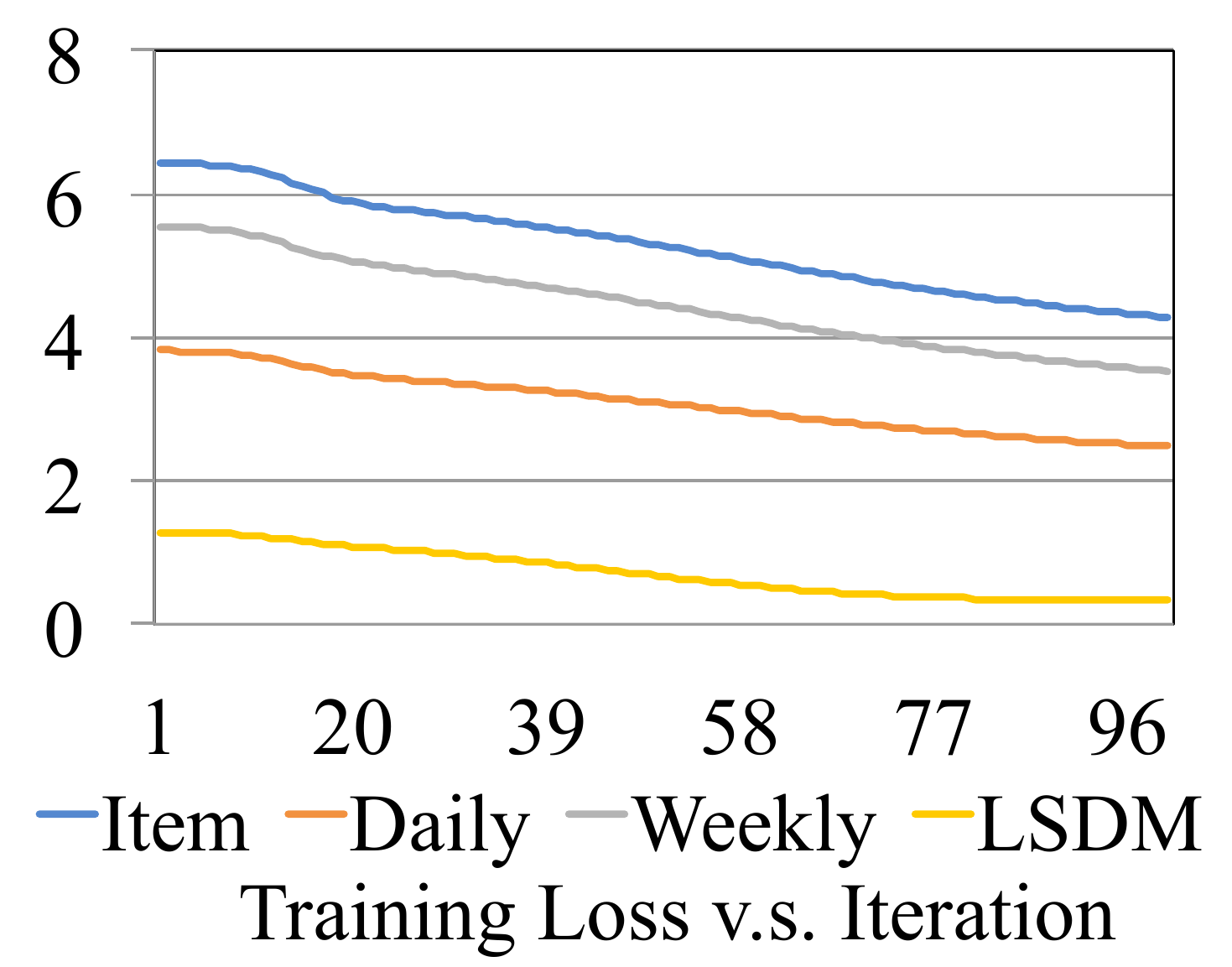}
       \includegraphics[width=1.53in]{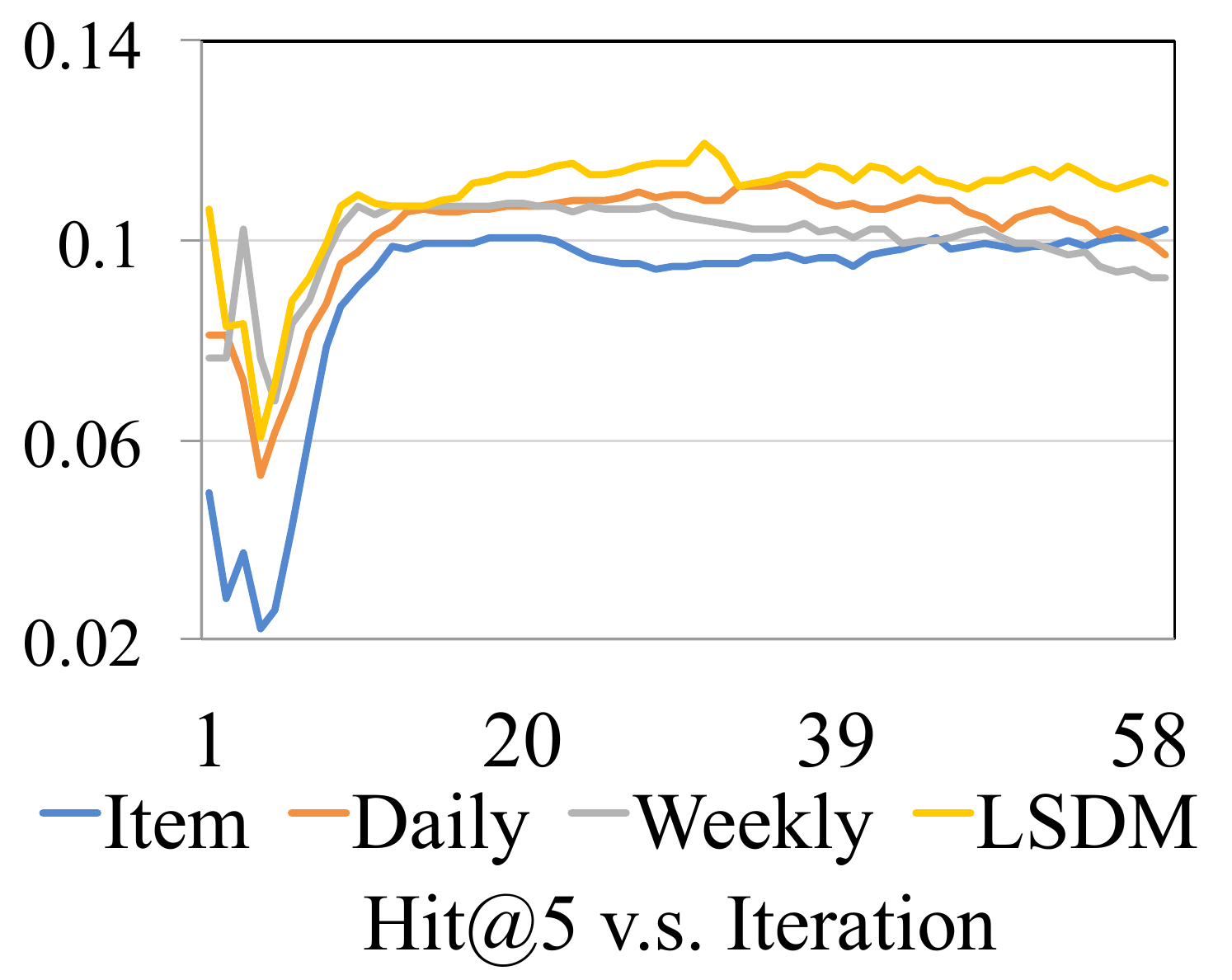}
     }
      \subfigure[BeiRen]{
     \includegraphics[width=1.53in]{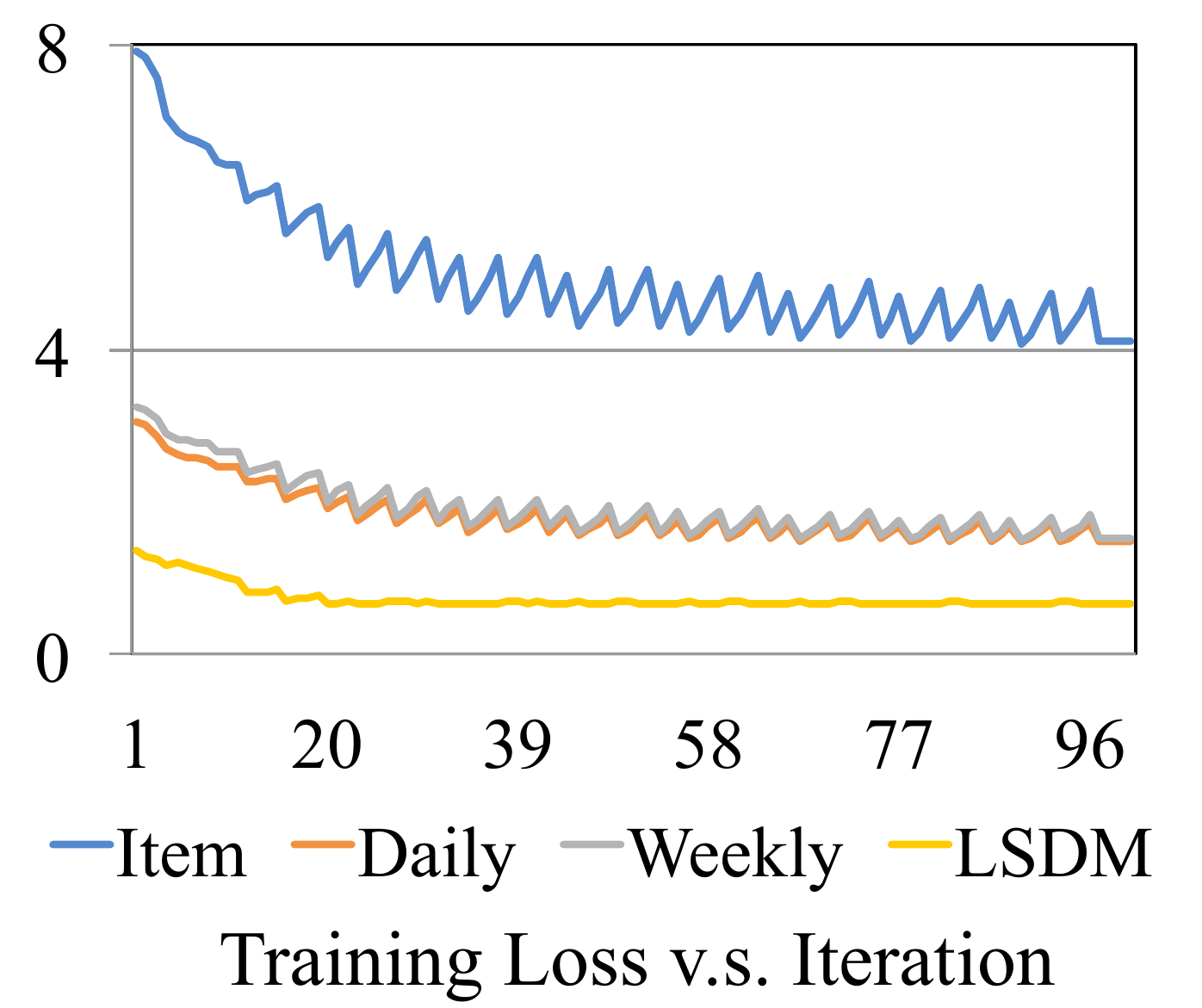}
    \includegraphics[width=1.53in]{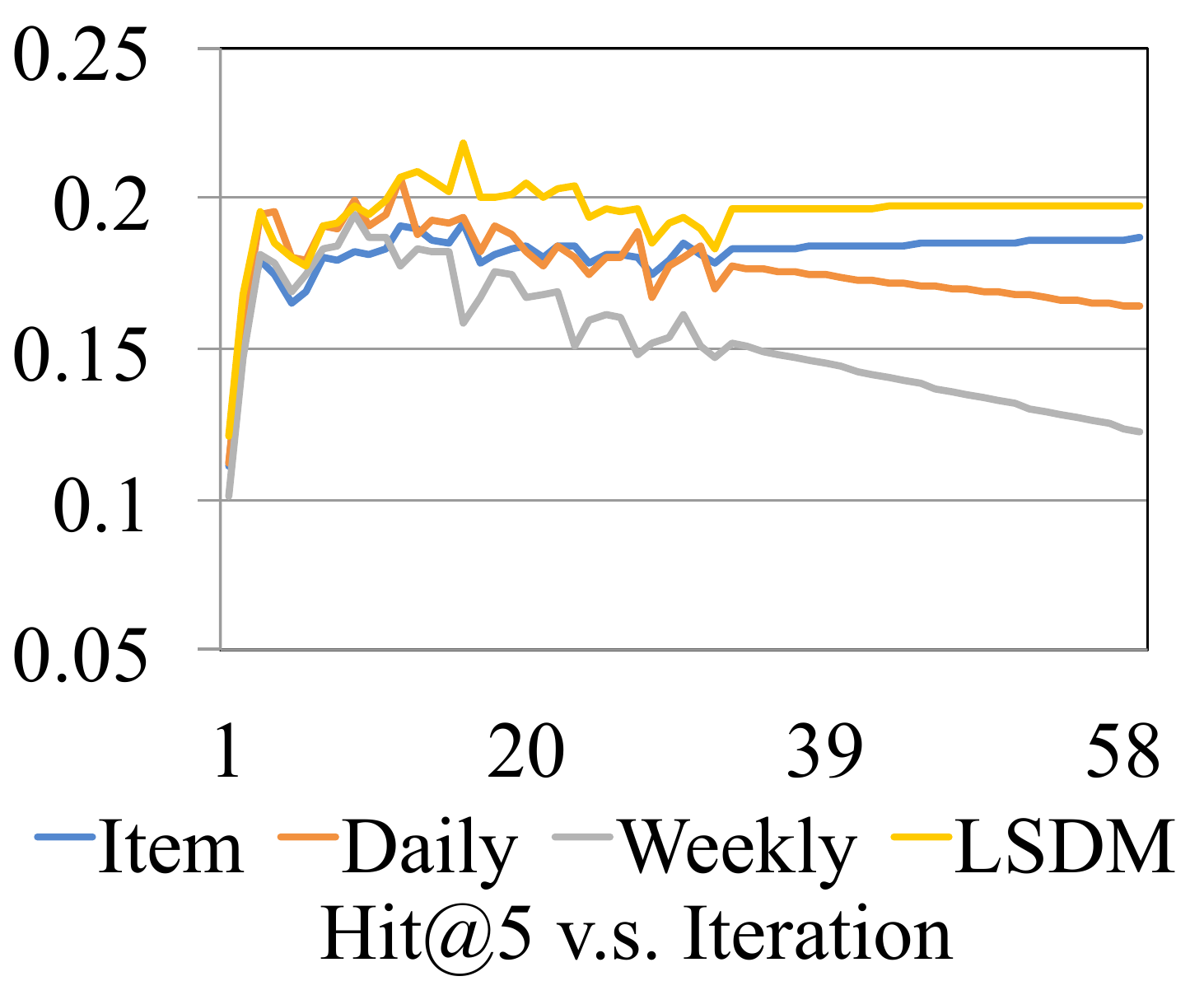}
    }
  \caption{Training loss and recommendation performance on Hit@5 of LDSM and degenerated models with different time scales w.r.t. the number of iterations on Ta-Feng and BeiRen datasets.}
  \label{fig-training} 
\end{figure*}}

\begin{figure*}
  \centering
  \subfigure[Ta-Feng: Hit@5]{
    \label{fig:subfig:b} 
     \includegraphics[width=2.1in]{figures/hit-tafeng.pdf}}
  \subfigure[Ta-Feng: NDCG@5]{
    \includegraphics[width=2.1in]{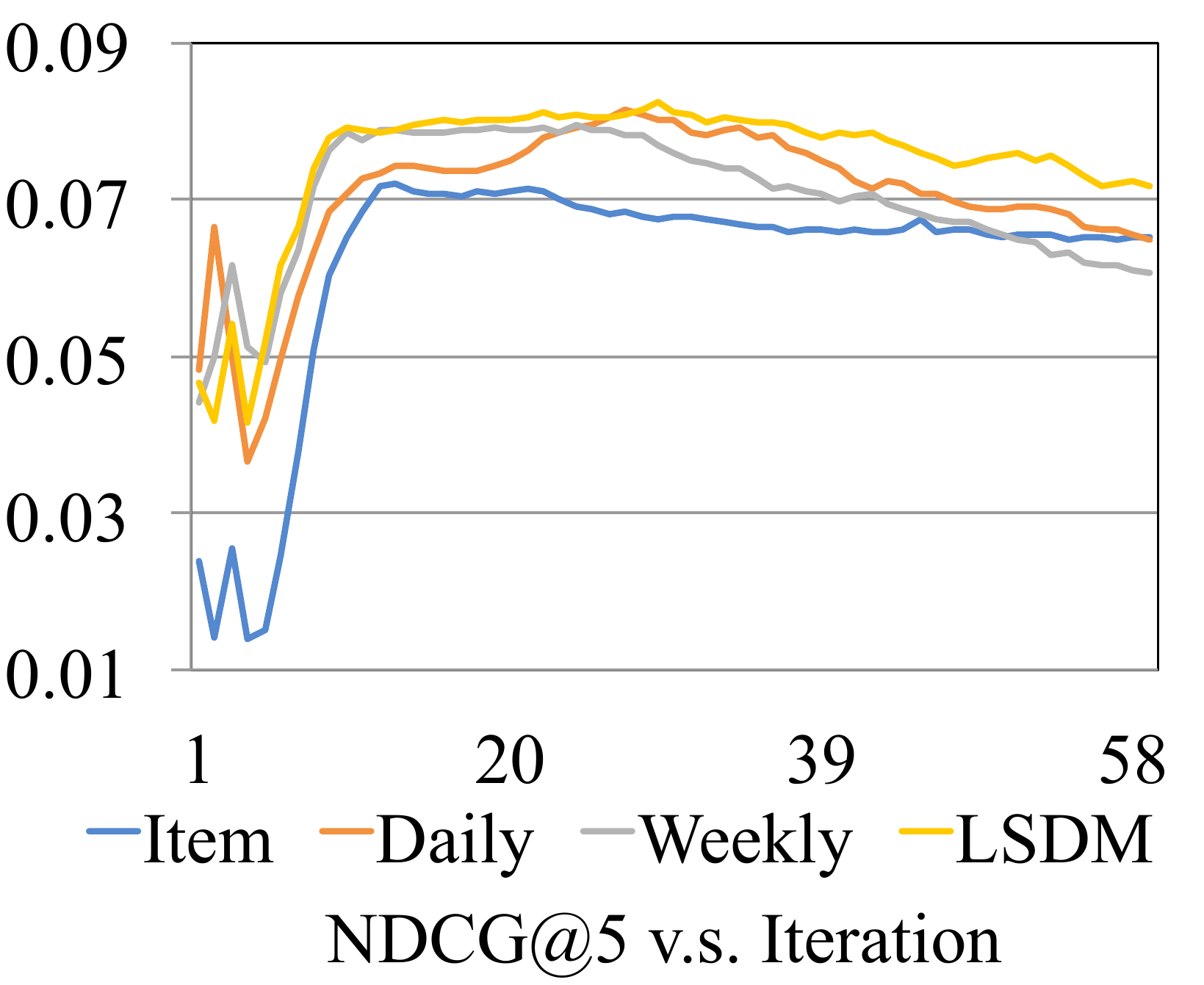}}
\subfigure[Ta-Feng: Training Loss]{
    \includegraphics[width=2.2in]{figures/loss-tafeng.pdf}}
      \subfigure[BeiRen: Hit@5]{
     \includegraphics[width=2.2in]{figures/hit-beiren.pdf}}
  \subfigure[BeiRen: NDCG@5]{
    \includegraphics[width=2.2in]{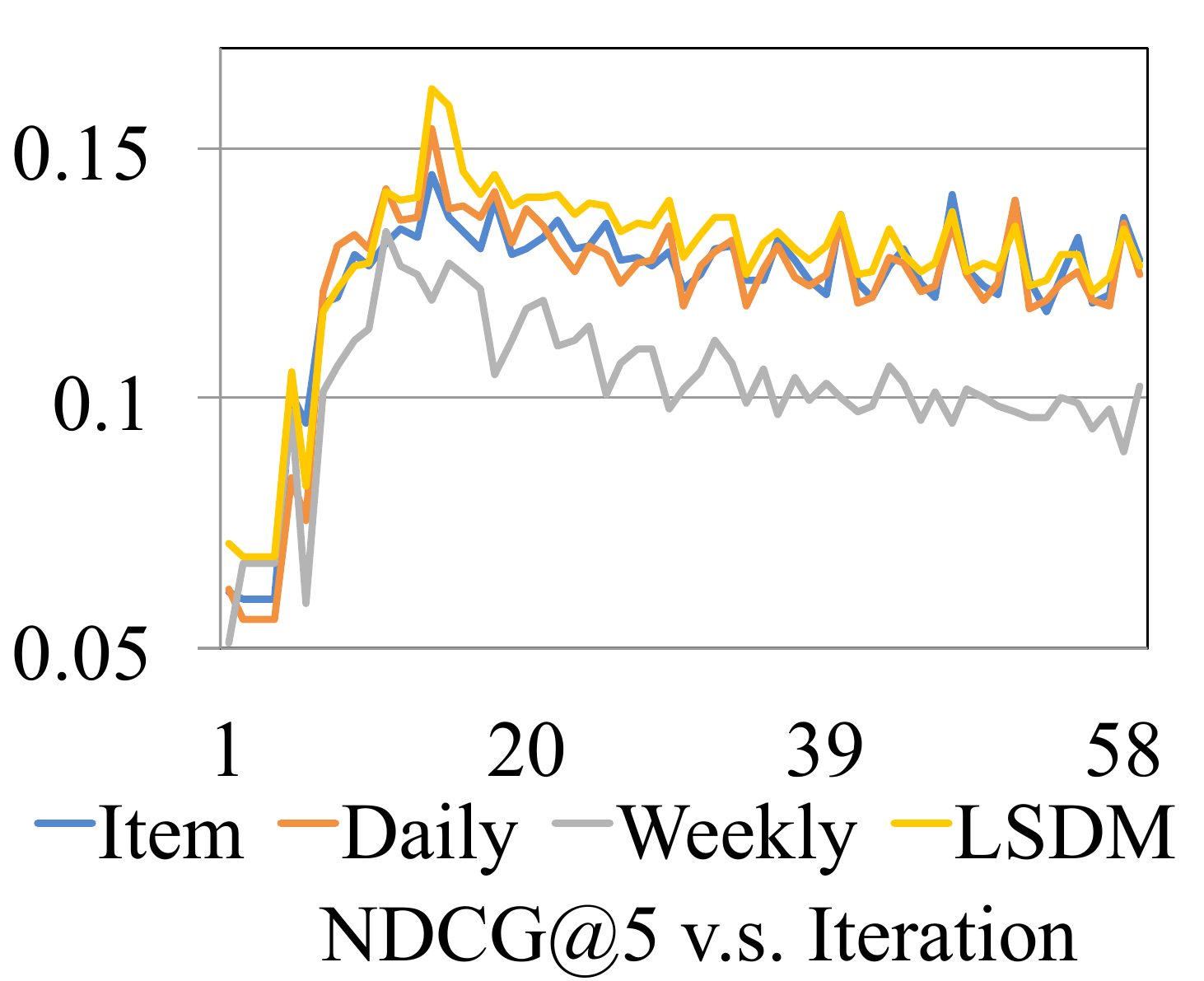}}
\subfigure[BeiRen: Training Loss]{
    \includegraphics[width=2.1in]{figures/loss-beiren.pdf}}
  \caption{Training loss and recommendation performance on Hit@5 and NDCG@5 of LSDM and degenerated models with different time scales w.r.t. the number of iterations on Tafeng and BeiRen datasets.}
  \label{fig-training} 
\end{figure*}

%% file: sec-related.tex
\section{Related Work}
Recommender systems have attracted a lot of attentions from the research community and industry. According to whether the sequence information is used, we summarizes the related methods of recommender system as follows.

\paratitle {Non-sequential Methods.}
Traditional non-sequential approaches at the early stage can be roughly divided into two categories, namely memory-based approaches and model-based approaches~\cite{sarwar2001item}.
Memory-based methods mainly rely on the neighborhood information for collaborative filtering; while model-based methods try to learn a prediction function using the history data. Model-based collaborative filtering methods such as matrix factorization algorithms and their variants have been proven to be effective to address the scalability and sparsity challenges in recommendation tasks~\cite{koren2009matrix, rendle2009bpr, Yehuda:FMN, chen2017gloma}.
Recently, deep learning techniques have been successfully applied in recommendation tasks and some pioneering studies have yielded promising results. Deep recommendation models mainly utilize deep learning techniques as a powerful data representation model, in which complicated user-item interactions and auxiliary information can be modeled in a unified representation. For example, neural rating prediction~\cite{Ruslan:RBM}, neural collaborative filtering~\cite{NCF, bai2017neural} and auto-encoder based recommender~\cite{Sedhain:Autorec,Hao:CDL,wu2016collaborative}. 
Those traditional approaches and the neural methods do not utilize sequential information, which disable them to capture user's varying appetite of items over time.

\paratitle {Sequential Methods.}
Detecting the purchase appetites of users and their evolution over time has been an active research topic in recent years. The main approaches to model the sequential behavior of a user have been developed in different recommendation settings: next-basket recommendation~\cite{rendle2010factorizing,wang2015learning,yu2016dynamic,guidotti2017next}, session-based~\cite{hidasi2015session,hidasi2016parallel,quadrana2017personalizing,li2017neural,jannach2017recurrent} and direct transaction-based recommendation~\cite{song2016multi,he2017translation,donkers2017sequential,wu2017recurrent,beutel2018latent,wu2018neural}.
The next-basket prediction aims at predicting what items the user could put in his basket.
The items are certainly dependent on the general interests of the user, but are also dependent on the items that the user has purchased in both his previous baskets and current basket. Two main approaches have been used to address the next basket recommendation problem:  Markov Chains (MC) and Recurrent Neural Network (RNN). The Factorizing Personalized Markov Chains (FPMC) approach~\cite{rendle2010factorizing} models both user's sequential behavior and general tastes by conducting a tensor factorization over the transition cubes. The RNN based model, \eg Hierarchical Representation Model (HRM)~\cite{wang2015learning}, improves FPMC by employing a two-layer architecture to construct a non-linear hybrid aggregation of the user profile vector and the transaction representation. Dynamic REcurrent bAasket Model~\cite{yu2016dynamic} (DREAM) adopts RNN to model global sequential features which reflect interactions among baskets, and uses the hidden state of RNN to represent user's dynamic interests over time. 
Session-based sequence models are commonly used in the web page clicking scenarios. It is different from next basket recommendation in that the order of clicks on items in a session is  considered.  RNN-based methods~\cite{medsker2001recurrent,hidasi2015session,hidasi2016parallel,quadrana2017personalizing,li2017neural,jannach2017recurrent} are usually adopted to capture the long historical records of users.
In ~\cite{quadrana2017personalizing}, the user's characteristics are learned by modeling user's representation in the sequence. The rich features of items are also incorporated into RNN model to learn the preference of users~\cite{hidasi2016parallel}. To make more accurate prediction,  attention mechanism is utilized in~\cite{li2017neural} to capture user's main interests in the current session.
Different from basket and session-based methods, which generally cluster items explicitly into baskets or sessions, some transaction-based approaches directly model the sequence of transaction of items~\cite{song2016multi,he2017translation,donkers2017sequential,wu2017recurrent,beutel2018latent,wu2018neural}. In addition, sequential patterns have also been extracted to  reflect the co-occurrences (or dependencies) of items or periodical characteristic of item purchases~\cite{mobasher2002using,tzvetkov2005tsp,yap2012effective,guidotti2017next}. Recent work also leverage low-rank tensor completion and product category inter-purchase duration vector~\cite{yi2017demand} to model the duration of items instead of the time scales of purchases. 

In all the above studies, we observe that the models work on a single sequence of items, transactions or sessions. The previous studies demonstrated that some types of purchase patterns can be extracted from such a sequence, but none of them attempted to extract different patterns at different time scales. This latter is exactly the goal of our study - our LSDM considers several sequences at different time scales so as to draw a more complete picture of the sequential behavior of the user and allows us to discover various co-purchase patterns and repeated purchasing at different time scales. Modeling users' purchase with multi-time scales enables our model to better understand the real-time purchase demands of users and recommend the items at the right time. We will show in our experiments that this results in better predictions.

%% file: sec-conclusion.tex
\section{Conclusion}
In this paper, we explored the utilization of different time scales for next-item recommendation. Our assumption was that different long- and short time purchase demands (\ie repetitive purchase and co-purchase) of users can exhibit with different time scales. This assumption was validated by the experimental results in our model on next-item recommendation task.
Our proposed Long-Short Demands-aware Model (LSDM) captures both user's interests towards items and user's demands over time. Experimental results on three public datasets (\ie Ta-Feng, BeiRen and Amazon) demonstrate the effectiveness of our model. 
While the idea of using multiple time scales is validated, our implementation can be further improved, with respect to detect the best time scales from the data automatically. It is also possible to incorporate richer information in the recommendation process, such as attribute information of items (\ie category, price) and textual description of items, etc. We will explore these avenues in the future.